\documentclass[twocolumn,conference]{IEEEtran}
\usepackage[T1]{fontenc}
\usepackage[latin9]{inputenc}
\usepackage{float}
\usepackage{calc}
\usepackage{amsmath}
\usepackage{amssymb}
\usepackage{graphicx}
\usepackage{esint}
\usepackage[unicode=true,
 bookmarks=true,bookmarksnumbered=true,bookmarksopen=true,bookmarksopenlevel=1,
 breaklinks=false,pdfborder={0 0 0},pdfborderstyle={},backref=false,colorlinks=false]
 {hyperref}
\hypersetup{pdftitle={Your Title},
 pdfauthor={Your Name},
 pdfpagelayout=OneColumn, pdfnewwindow=true, pdfstartview=XYZ, plainpages=false}

\makeatletter

\providecommand{\tabularnewline}{\\}

\usepackage{subfig}

\usepackage{balance}

\@ifundefined{showcaptionsetup}{}{%
 \PassOptionsToPackage{caption=false}{subfig}}
\usepackage{subfig}
\makeatother

\begin{document}
\setlength{\abovedisplayskip}{0pt}

\setlength{\belowdisplayskip}{0pt}
\title{Multilevel Polar Coded Space-Shift Keying}
\author{\IEEEauthorblockN{Draft~1\IEEEauthorrefmark{1}}\IEEEauthorblockA{School of Electrical and\\
Computer Engineering\\
Institute of Technology\\
99999 Testcity\\
Email: test@test.tes}\and \IEEEauthorblockN{Second~Name}\IEEEauthorblockA{Ecole Superieure\\
Nantes, France\\
Email: second@second.fr}\and \IEEEauthorblockN{Third~Name\\
and Fourth~Name}\IEEEauthorblockA{Star Academy\\
San Francisco, California 99999-9999\\
Telephone: (800) 555\textendash 5555\\
Fax: (888) 555\textendash 5555}}
\author{\IEEEauthorblockN{Muhammad~Zaeem~Hasan, Nemanja~Stefan~Perovi\'c and Mark~F.~Flanagan
}\IEEEauthorblockA{School of Electrical and Electronic Engineering, University College
Dublin, Belfield, Dublin 4, Ireland\\
Email: muhammad.hasan@ucdconnect.ie, nemanja.stefan.perovic@ucd.ie,
mark.flanagan@ieee.org}}
\IEEEspecialpapernotice{}
\IEEEaftertitletext{after title text like dedication}
\maketitle
\begin{abstract}
Multilevel coding (MLC) is a coded modulation technique which can
achieve excellent performance over a range of communication channels.
 Polar codes have been shown to be quite compatible with communication
systems using MLC, as the rate allocation of the component polar codes
follows the natural polarization inherent in polar codes. MLC based
techniques have not yet been studied in systems that use spatial modulation
(SM). SM makes the polar code design difficult as the spatial bits
actually select a channel index for transmission. To solve this problem,
we propose a Monte Carlo based evaluation of the ergodic capacities
for the individual bit levels under the capacity rule for a space-shift
keying (SSK) system, where we also make use of a single antenna activation
to approximate the transmission channel for the design of the multilevel
polar code. Our simulation results show that the multilevel polar
coded $16\times1$ SSK system outperforms the corresponding system
that uses bit-interleaved polar coded modulation by $2.9$ dB at a
bit-error rate (BER) of $10^{-4}$.
\end{abstract}


\IEEEpeerreviewmaketitle{}

\section{Introduction}

Multilevel coded (MLC) modulation was first introduced independently
by Imai and Hirakawa \cite{j2}, and Ungerboeck \cite{j1},\cite{j3}.
MLC exhibits a performance gain over bit-interleaved coded modulation
(BICM) by considering interdependency of the bits that map to a constellation
symbol. On the other hand, in BICM \cite{j1a}, the interleaver removes
any dependency among the adjacent bits, and thus helps in simplifying
the receiver design, however at a cost of decreased performance compared
to MLC.

Polar codes have been shown to have excellent performance when used
with the MLC design paradigm. The polarization effect in a larger
polar code has been proved to be equivalent to that of concatenating
smaller polar codes that constitute the larger polar code. In other
words, rate allocation of the component polar codes using the capacity
rule \cite{j4} is the same as designing a larger polar code and then
dividing it into polar codes having different rates, as proved in
\cite{j5}. In \cite{j5}, the authors have described that the rate
allocation in polar codes follows the capacity rule when multi-stage
decoding is performed, which is equivalent to successive cancellation.
However, the problem of multilevel polar code design has been studied
only for standard constellations such as amplitude-shift keying (ASK),
phase-shift keying (PSK) and quadrature amplitude modulation (QAM).
Multilevel coded modulation and polar code design has yet to be applied
to multiple-antenna index modulation schemes such as space-shift keying
(SSK) and spatial modulation (SM).

SM was developed as an alternative to space-time and spatial-multiplexing
techniques for the multi-input multi-output (MIMO) channel \cite{j4a}.
SM maps one part of the information bits to select a particular antenna
for transmission and the others to choose a constellation symbol for
transmission from that antenna. Due to the single antenna transmission
in SM, the receiver design becomes simple as there is no inter-symbol
or inter-channel interference \cite{j4b}. Space-shift keying (SSK)
modulation is a special case of SM where the information is transmitted
by using only the antenna index \cite{j4c}. As there is no need to
detect the constellation symbol in SSK, the receiver complexity is
further reduced. 

In this paper, we have designed the rates of multilevel polar codes
using the capacity rule for multilevel SSK modulation. This is achieved
by using the Monte-Carlo method to evaluate the ergodic capacities
of the different bit levels of SSK modulation. The different bit levels
of SSK are shown to have quite different bit-level capacities, which
further motivates our approach. For the sake of simplicity of the
system design, we have designed the polar code for an average-case
scenario of the Rayleigh fading channel using the method given in
\cite{j13}. Our simulation results show that at a bit error rate
(BER) of $10^{-4}$, the designed MLC polar coded $16\times1$ SSK
system exhibits a gain of $2.9$ dB over the corresponding system
using BICM.

The rest of the paper is structured as follows. We first present the
system model in Section \ref{sec:SystemModel}. In Section \ref{sec:Proposed-Method},
we show how to compute the ergodic capacities of the different bit
levels in SSK modulation using the capacity rule, and we present the
respective polar code design. In Section \ref{sec:Results-and-Discussion},
we present numerical results and we conclude our paper in Section
\ref{sec:Conclusion}.

\begin{figure*}[tp]
\begin{centering}
\includegraphics[width=1\textwidth]{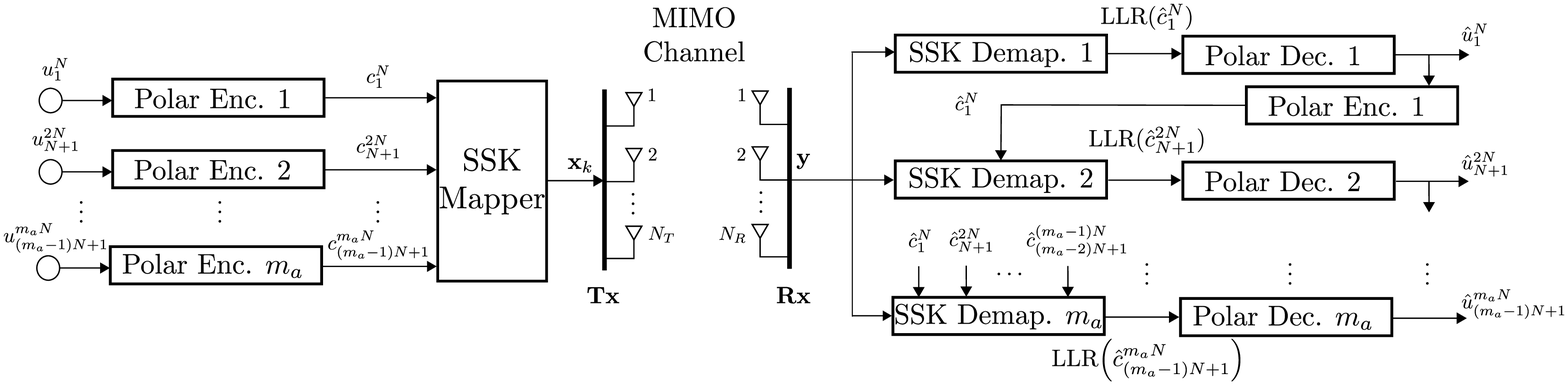}
\par\end{centering}
\caption{\label{Fig1}Proposed multilevel polar coded SSK system, with two-stage
polar transform at the transmitter and corresponding multi-stage receiver.}
\end{figure*}

\section{System Model\label{sec:SystemModel}}

\subsection{Multilevel Polar Code}

The proposed multilevel polar coded SSK system is illustrated in Fig.
\ref{Fig1}. A message block of $K=\sum_{i=1}^{m_{a}}K_{i}$ bits
is divided into $m_{a}$ modulation streams, where the $i$th polar
encoder encodes $K_{i}$ bits using a polar code of rate $R_{i}=K_{i}/N$
and length $N=2^{n}$, $n>0$. At the $i$th encoding level of the
transmitter, information bits $u_{\mathcal{A_{\mathrm{\mathit{i}}}}}$
and frozen bits $u_{\mathcal{A_{\mathrm{\mathit{i}}}^{\mathrm{\mathit{c}}}}}$,
where $\left|\mathcal{A}_{i}\right|=K_{i}$ and $\left|\mathcal{A}_{\mathrm{\mathit{i}}}^{\mathrm{\mathit{c}}}\right|=N-K_{i}$,
are combined in an uncoded vector $u_{(i-1)N+1}^{iN}=\left[u_{(i-1)N+1}u_{(i-1)N+2}\ldots u_{iN}\right]$
and polar coded to form a codeword $c_{(i-1)N+1}^{iN}$ which is sent
to the SSK modulator. The total code rate of the polar coded system
is $R=K/(m_{a}N)$.

\subsection{SSK Modulation and Channel Transmission}

As the SSK modulated symbol selects a particular antenna $k\in\left\{ 1,2,\ldots,N_{T}\right\} $
for transmission, the received signal for the MIMO system with $N_{T}=2^{m_{a}}$
transmit and $N_{R}$ receive antennas can be written as

\begin{equation}
\mathbf{y}=\mathbf{h}_{k}+\mathbf{n},\label{eq:1}
\end{equation}
where $\mathbf{h}_{k}$ is the $k$th column of the $N_{R}\times N_{T}$
complex Rayleigh fading channel matrix $\mathbf{H}$ with independent
and identically distributed (i.i.d.) coefficients \emph{$h_{pq}\sim\mathcal{CN}(0,1)$,
}where $p\in\left\{ 1,\ldots,N_{R}\right\} ,\:q\in\left\{ 1,\ldots,N_{T}\right\} $,
\emph{$\mathbf{n}\sim\mathcal{CN}(\boldsymbol{0},N_{0}\mathbf{I})$
}is a $N_{R}\times1$ vector containing independent zero-mean circularly-symmetric
complex Gaussian entries, each with variance $N_{0}$, where $N_{0}$
is the noise power spectral density and $\mathbf{I}$ is the $N_{R}\times N_{R}$
identity matrix, and \emph{$\mathbf{y}$ }is the $N_{R}\times1$ complex
received signal vector. The number of bits required to represent a
SSK symbol is $m_{a}=\log_{2}(N_{T})$.

\subsection{Multi-Stage Decoder}

At the receiver, successive detection of the bits at the different
SSK modulation levels is performed by using a series of multi-stage
soft demappers and successive cancellation (SC) polar decoders, as
shown in Fig. \ref{Fig1}. At the $i$th decoding level, the SSK soft
demapper takes the output from all of the previous decoding stages
and provides log-likelihood ratios (LLRs) to the SC polar decoder
which in turn produces the estimated uncoded bits $\hat{u}_{(i-1)N+1}^{iN}$
for that particular stage. In the next step, these bits are polar
encoded to form the codeword estimate $\hat{c}_{(i-1)N+1}^{iN}$,
and this is fed to the SSK demapper of the next stage for assistance
in forming the soft bit estimates. This process is repeated until
the the final ($m_{a}$th) stage, where the decoding results from
all previous stages are used to inform the decoding of the current
stage.

\section{Multilevel Polar Code Design for SSK\label{sec:Proposed-Method}}

In order to design the rates of the multi-level polar code, the bit-level
capacities of MLC SSK modulation need to be computed. The bit-level
capacities can be evaluated using a method similar to that given in
\cite{j4} with the difference that here the modulated symbols actually
represent different selected channels instead of different constellation
symbols. Another difference is that the bit-level capacities need
to be averaged over the fading channel statistics in order to provide
an accurate representation of the average-case rates; this is elucidated
in the Subsections \ref{subsec:SSK-MLC} and \ref{subsec:Polar-Code-Rate}.
As an illustration we evaluate the 4 multi-level ergodic capacities
for 16-SSK in Subsection \ref{subsec:16-SSK-Bit-Level-Capacities}.
Finally, Subsection \ref{subsec:Polar-Code-Rate} describes the rate
allocation method for the multi-level polar code.

\subsection{SSK Multi-Level Coding Capacity\label{subsec:SSK-MLC}}

\begin{figure}[H]
\centering{}\includegraphics[scale=0.7]{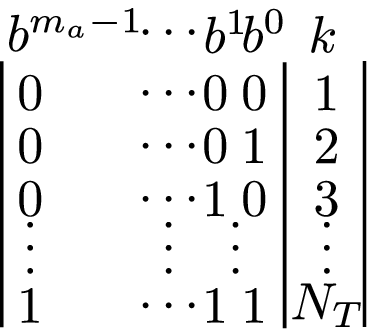}\caption{\label{Fig2}Binary representation of the SSK symbols for $N_{T}$
transmit antennas.}
\end{figure}

Let the symbols to represent $N_{T}$ transmit antennas be equiprobable
in the set $\mathcal{X}=\left\{ 1,2,\ldots,N_{T}\right\} $ with the
corresponding total SSK capacity $C(\mathcal{X})$ given as

\begin{equation}
\begin{aligned}C(\mathcal{X}) & =\frac{1}{N_{T}}\intop_{\mathbf{y}}\sum_{k=1}^{N_{T}}p(\mathbf{y}\arrowvert k)\log_{2}\negthickspace\left(\frac{p(\mathbf{y}\arrowvert k)}{{\displaystyle \frac{1}{N_{T}}\sum_{k'=1}^{N_{T}}}p(\mathbf{y}\arrowvert k')}\right)d\mathbf{y},\end{aligned}
\label{eq:2}
\end{equation}
where the conditional probability density function of $\mathbf{y}$
given the antenna index $k$ is

\begin{equation}
p(\mathbf{y}\arrowvert k)=\frac{1}{(\pi N_{0})^{N_{R}}}e^{-\left\Vert \mathbf{y}-\mathbf{h}_{k}\right\Vert ^{2}/N_{0}}.\label{eq:3}
\end{equation}

We define the binary-to-decimal mapping $M$ by $M\left(\mathbf{b}^{m_{a}}\right)\!=\!k\in\mathcal{X}$,
where $\mathbf{b}^{m_{a}}=\left[b^{m_{a}-1}\ldots b^{1}b^{0}\right]$
and $b^{i}\in\left\{ 0,1\right\} $ for each $i$, as depicted in
the Fig.\ref{Fig2}. The inverse of $M$ is denoted by $M^{-1}\left(k\right)\!=\!\mathbf{b}^{m_{a}}$.
The capacity of the symbol subset $\mathcal{X}_{\textrm{\ensuremath{\mathbf{b}^{i}}}}=\left\{ k=M\left(\left[q^{m_{a}-1}\ldots q^{i}b^{i-1}\ldots b^{1}b^{0}\right]\right)\right\} $,
where $i\leq m_{a}$, $q^{i}\in\left\{ 0,1\right\} $ and $\mathbf{b}^{i}=\left[b^{i-1}\ldots b^{1}b^{0}\right]$,
can be written as

\begin{equation}
C_{i}^{0}=\frac{1}{2^{i}}\sum_{b^{0}=0}^{1}\negthickspace\ldots\negthickspace\sum_{b^{i-1}=0}^{1}\intop_{\mathbf{y}}p(\mathbf{y}\arrowvert\mathbf{b}^{i})\log_{2}\negthickspace\left(\frac{p(\mathbf{y}\arrowvert\mathbf{b}^{i})}{p(\mathbf{y})}\right)d\mathbf{y},\label{eq:4}
\end{equation}
where

\begin{equation}
\begin{aligned}p(\mathbf{y}\arrowvert\mathbf{b}^{i}) & =\frac{2^{i}}{N_{T}}\sum_{k\in\mathcal{X}_{\textrm{\ensuremath{\mathbf{b}^{i}}}}}\negthickspace p(\mathbf{y}\arrowvert k).\end{aligned}
\label{eq:5}
\end{equation}

Using the method given in \cite{j4}, we can obtain the capacity of
the $i$th bit level as

\begin{equation}
\begin{aligned}C^{i} & =\begin{cases}
C_{i}^{0}, & i=1,\\
C_{i}^{0}-C_{i-1}^{0}, & 2\leq i\leq m_{a}-1,\\
C\left(\mathcal{X}\right)-C_{i-1}^{0}, & i=m_{a}.
\end{cases}\end{aligned}
\label{eq:6}
\end{equation}

Note that the spatial bits choose a particular transmit antenna and
therefore a particular random channel. Therefore, we need to take
expectation over the channel statistics in order to obtain ergodic
bit-level capacities, as explained in the next subsection.

\subsection{Illustrative Example: Bit-Level Ergodic Capacities for \mbox{16-SSK}\label{subsec:16-SSK-Bit-Level-Capacities}}

In order to illustrate our approach, in the following we provide an
example which shows how to calculate the bit-level ergodic capacities
for the case of 16-SSK modulation.

\subsubsection{Bit-Level-1 Ergodic Capacity\label{subsec:Bit-Level-1-Capacity}}

Using \eqref{eq:4} and \eqref{eq:6}, we define the capacity of bit-level
$i=1$, where $\mathbf{b}^{i}=b^{0}$, as

\begin{equation}
\begin{aligned}C_{1}^{0} & =\frac{1}{2^{1}}\sum_{b^{0}=0}^{1}\intop_{\mathbf{y}}p(\mathbf{y}\arrowvert\mathbf{b}^{i})\log_{2}\negthickspace\left(\frac{p(\mathbf{y}\arrowvert\mathbf{b}^{i})}{p(\mathbf{y})}\right)d\mathbf{y},\\
 & =\frac{1}{2}\sum_{j=0}^{2^{1}-1}C_{1,j},
\end{aligned}
\label{eq:7}
\end{equation}
where

\begin{equation}
\begin{aligned}C_{1,j} & =\intop_{\mathbf{y}}p(\mathbf{y}\arrowvert\mathbf{b}^{i}\!=\!M^{-1}\left(j\right))\log_{2}\negthickspace\left(\frac{p\left(\mathbf{y}\arrowvert\mathbf{b}^{i}\!=\!M^{-1}\left(j\right)\right)}{p(\mathbf{y})}\right)d\mathbf{y}.\end{aligned}
\label{eq:8}
\end{equation}
Next we define the symbol subset $\mathcal{X}_{\mathbf{b}^{i}=M^{-1}\left(j\right)}$,
which is the set of all values $M(\mathbf{b}^{i})$ where the initial
$i$ bits of the length $m_{a}$ vector are equal to $M^{-1}\left(j\right)$.
With this, we then expand \eqref{eq:7} and \eqref{eq:8} for $j=0$
as

\begin{equation}
\begin{aligned}C_{1,0} & =\intop_{\mathbf{y}}p(\mathbf{y}\arrowvert\mathbf{b}^{i}\!=\!M^{-1}\left(0\right))\log_{2}\negthickspace\left(\negthickspace\frac{p\left(\mathbf{y}\arrowvert\mathbf{b}^{i}\!=\!M^{-1}\left(0\right)\right)}{p(\mathbf{y})}\negthickspace\right)\negthickspace d\mathbf{y},\\
 & =\intop_{\mathbf{y}}\frac{2}{N_{T}}\sum_{k\in\mathcal{X}_{\mathbf{b}^{i}=M^{-1}\left(0\right)}}\negthickspace\negthickspace\negthickspace\negthickspace\negthickspace\negthickspace\negthickspace p(\mathbf{y}\arrowvert k)\log_{2}\negthickspace\left(\negthickspace\frac{\frac{2}{N_{T}}\negthickspace{\displaystyle \sum_{k\in\mathcal{X}_{\mathbf{b}^{i}=M^{-1}\left(0\right)}}\!\negthickspace\negthickspace\negthickspace\negthickspace\negthickspace\negthickspace}p(\mathbf{y}\arrowvert k)}{\frac{1}{N_{T}}{\displaystyle \sum_{k'=1}^{N_{T}}}p(\mathbf{y}\arrowvert k')}\negthickspace\right)\negthickspace d\mathbf{y},\\
 & =\negthickspace\frac{2}{N_{T}}\negthickspace\intop_{\mathbf{y}}\!\negthickspace\sum_{k\in\mathcal{X}_{\mathbf{b}^{i}=M^{-1}\left(0\right)}}\negthickspace\negthickspace\negthickspace\negthickspace\negthickspace\negthickspace\negthickspace p(\mathbf{y}\arrowvert k)\log_{2}\negthickspace\left(\frac{2\negthickspace\negthickspace\negthickspace{\displaystyle \sum_{k\in\mathcal{X}_{\mathbf{b}^{i}=M^{-1}\left(0\right)}}}\!\negthickspace\negthickspace\negthickspace\negthickspace\negthickspace\negthickspace e^{-\left\Vert \mathbf{y}-\mathbf{h}_{k}\right\Vert ^{2}/N_{0}}}{{\displaystyle \sum_{k'=1}^{N_{T}}}e^{-\left\Vert \mathbf{y}-\mathbf{h}_{k'}\right\Vert ^{2}/N_{0}}}\negthickspace\right)\negthickspace d\mathbf{y}\!,\\
 & ={\displaystyle \underset{\mathbf{y\arrowvert}k\in\mathcal{X}_{\mathbf{b}^{i}=M^{-1}\left(0\right)}}{\mathbb{E}}}\left\{ \log_{2}\negthickspace\left(\frac{2\negthickspace\negthickspace\negthickspace{\displaystyle \sum_{k\in\mathcal{X}_{\mathbf{b}^{i}=M^{-1}\left(0\right)}}\!\negthickspace\negthickspace\negthickspace\negthickspace\negthickspace}\negthickspace e^{-\left\Vert \mathbf{y}-\mathbf{h}_{k}\right\Vert ^{2}/N_{0}}}{{\displaystyle \sum_{k'=1}^{N_{T}}}e^{-\left\Vert \mathbf{y}-\mathbf{h}_{k'}\right\Vert ^{2}/N_{0}}}\negthickspace\right)\right\} ,
\end{aligned}
\label{eq:9}
\end{equation}
where, $\mathbb{E}(\cdot)$ is the expectation operator and $\mathbf{b}^{i}=M^{-1}\left(0\right)$
implies $b^{0}=0$. Similarly for $j=1$ we have

\begin{equation}
\begin{aligned}C_{1,1} & =\intop_{\mathbf{y}}p(\mathbf{y}\arrowvert\mathbf{b}^{i}\!=\!M^{-1}\left(1\right))\log_{2}\left(\frac{p\left(\mathbf{y}\arrowvert\mathbf{b}^{i}\!=\!M^{-1}\left(1\right)\right)}{p(\mathbf{y})}\right)d\mathbf{y},\\
 & ={\displaystyle \underset{\mathbf{y\arrowvert}\mathcal{X}_{\mathbf{b}^{i}=M^{-1}\left(1\right)}}{\mathbb{E}}}\left\{ \log_{2}\left(\frac{2\negthickspace{\displaystyle \sum_{k\in\mathcal{X}_{\mathbf{b}^{i}=M^{-1}\left(1\right)}}\!\negthickspace\negthickspace\negthickspace\negthickspace\negthickspace\negthickspace}e^{-\left\Vert \mathbf{y}-\mathbf{h}_{k}\right\Vert ^{2}/N_{0}}}{{\displaystyle \sum_{k'=1}^{N_{T}}}e^{-\left\Vert \mathbf{y}-\mathbf{h}_{k'}\right\Vert ^{2}/N_{0}}}\right)\right\} .
\end{aligned}
\label{eq:10}
\end{equation}

Using \eqref{eq:6} and \eqref{eq:7} we can find the bit-level-1
capacity as $\begin{aligned}C^{1} & =C_{1}^{0}.\end{aligned}
$

\subsubsection{Bit-Level-2 Ergodic Capacity\label{subsec:Bit-Level-2-Capacity}}

Using \eqref{eq:4} and \eqref{eq:6}, we define the capacity of bit-level
$i=2$ , where $\mathbf{b}^{i}=b^{1}b^{0}$, as

\begin{equation}
\begin{aligned}C_{2}^{0} & =\frac{1}{2^{2}}\sum_{b^{1}=0}^{1}\sum_{b^{0}=0}^{1}\intop_{\mathbf{y}}p(\mathbf{y}\arrowvert\mathbf{b}^{i})\log_{2}\left(\frac{p\left(\mathbf{y}\arrowvert\mathbf{b}^{i}\right)}{p(\mathbf{y})}\right)d\mathbf{y},\\
 & =\frac{1}{4}\sum_{j=0}^{2^{2}-1}C_{2,j},
\end{aligned}
\label{eq:11}
\end{equation}
where

\begin{equation}
\begin{aligned}C_{2,j} & ={\displaystyle \underset{\mathbf{y\arrowvert}k\in\mathcal{X}_{\mathbf{b}^{i}\!=\!M^{-1}\left(j\right)}}{\mathbb{E}}}\left\{ \log_{2}\left(\frac{4\negthickspace\negthickspace{\displaystyle \sum_{k\in\mathcal{X}_{\mathbf{b}^{i}\!=\!M^{-1}\left(j\right)}}}\negthickspace\negthickspace\negthickspace\negthickspace\negthickspace e^{-\left\Vert \mathbf{y}-\mathbf{h}_{k}\right\Vert ^{2}/N_{0}}}{{\displaystyle \sum_{k'=1}^{N_{T}}}e^{-\left\Vert \mathbf{y}-\mathbf{h}_{k'}\right\Vert ^{2}/N_{0}}}\right)\right\} .\end{aligned}
\label{eq:12}
\end{equation}

Finally we can obtain the bit-level-2 capacity using \eqref{eq:6}
and \eqref{eq:11} as $\begin{aligned}C^{2} & =C_{2}^{0}-C_{1}^{0}.\end{aligned}
$

\subsubsection{Bit-Level-3 Ergodic Capacity\label{subsec:Bit-Level-3-Capacity}}

Using \eqref{eq:4} and \eqref{eq:6}, we define the capacity of bit-level
$i=3$, where $\mathbf{b}^{i}=b^{2}b^{1}b^{0}$, as

\begin{equation}
\begin{aligned}C_{3}^{0} & =\frac{1}{2^{3}}\sum_{b^{2}=0}^{1}\sum_{b^{1}=0}^{1}\sum_{b^{0}=0}^{1}\intop_{\mathbf{y}}p(\mathbf{y}\arrowvert\mathbf{b}^{i})\log_{2}\left(\frac{p\left(\mathbf{y}\arrowvert\mathbf{b}^{i}\right)}{p(\mathbf{y})}\right)d\mathbf{y},\\
 & =\frac{1}{8}\sum_{j=0}^{2^{3}-1}C_{3,j},
\end{aligned}
\label{eq:13}
\end{equation}
where

\begin{equation}
\begin{aligned}C_{3,j} & ={\displaystyle \underset{\mathbf{y\arrowvert}k\in\mathcal{X}_{\mathbf{b}^{i}\!=\!M^{-1}\left(j\right)}}{\mathbb{E}}}\left\{ \log_{2}\negthickspace\left(\frac{8\negthickspace\negthickspace{\displaystyle \sum_{k\in\mathcal{X}_{\mathbf{b}^{i}\!=\!M^{-1}\left(j\right)}}}\negthickspace\negthickspace\negthickspace\negthickspace\negthickspace\negthickspace e^{-\left\Vert \mathbf{y}-\mathbf{h}_{k}\right\Vert ^{2}/N_{0}}}{{\displaystyle \sum_{k'=1}^{N_{T}}}e^{-\left\Vert \mathbf{y}-\mathbf{h}_{k'}\right\Vert ^{2}/N_{0}}}\right)\right\} .\end{aligned}
\label{eq:14}
\end{equation}

We can obtain the bit-level-3 capacity by using \eqref{eq:6} and
\eqref{eq:13} as $\begin{aligned}C^{3} & =C_{3}^{0}-C_{2}^{0}.\end{aligned}
$

\subsubsection{Bit-Level-4 Capacity\label{subsec:Bit-Level-4-Capacity}}

The bit-level-4 capacity can be obtained using \eqref{eq:2} and \eqref{eq:13}
as $\begin{aligned}C^{4} & =C\left(\mathcal{X}\right)-C_{3}^{0}.\end{aligned}
$

\begin{figure}[tp]
\begin{centering}
\includegraphics[width=1\linewidth]{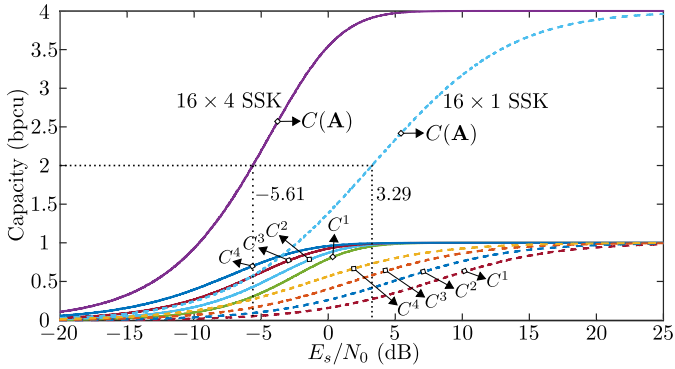}
\par\end{centering}
\caption{\label{Fig3}Different bit-level capacities for $16\times1$ and $16\times4$
SSK modulation.}

\end{figure}

\subsection{Polar Code Rate Allocation\label{subsec:Polar-Code-Rate}}

To design the rates for the different levels of the multilevel polar
code, we need to evaluate the overall SSK capacity and the corresponding
$m_{a}$ bit-level capacities as given in Section \ref{subsec:SSK-MLC}.
We have computed the bit-level capacities by evaluating the expectation
using Monte-Carlo simulation for a large number of frames for the
received signal $\mathbf{y}$. The expectation averages the effect
of the fast fading channels, resulting in the bit-level ergodic capacities.
As an example, Fig. \ref{Fig3} shows the different bit-level ergodic
capacities of $16\times1$ and $16\times4$ SSK modulations that are
found using the Monte-Carlo simulation.

The bit-level MLC capacities are chosen for a particular information
rate in bits per channel use (bpcu) of the overall SSK system using
the capacity rule, which states that the rate of the $i$th component
channel code should be $R_{i}\leq C^{i}$ for $1\leq i\leq m_{a}$.
This design targets the overall rate for the length $m_{a}N$ polar
code and the corresponding design signal-to-noise ratio (DSNR).

In the next step, we design the polar code of length $m_{a}N$ for
a SISO Rayleigh fading channel using Tal-Vardy's degrade transform
and degrade merge methods for the DSNR \cite{j8}, \cite{j13}. In
the case of SSK, the effective channel is SIMO rather than SISO, and
therefore this design is not capacity achieving, i.e., its capacity
is less than that of MIMO-SSK channel; however, it provides a simple
way to design the multilevel polar code for the SSK modulated system.

In the final step, we segregate the $m_{a}N$ length polar code into
$m_{a}$ cascaded component polar codes according to the rates found
using the MLC bit-level ergodic capacities as described in Section
\ref{subsec:SSK-MLC}.

\begin{figure*}[tp]
\begin{centering}
\subfloat[\label{fig:Fig4(a)} $16\times1$ SSK system]{\centering{}\includegraphics[width=0.5\textwidth]{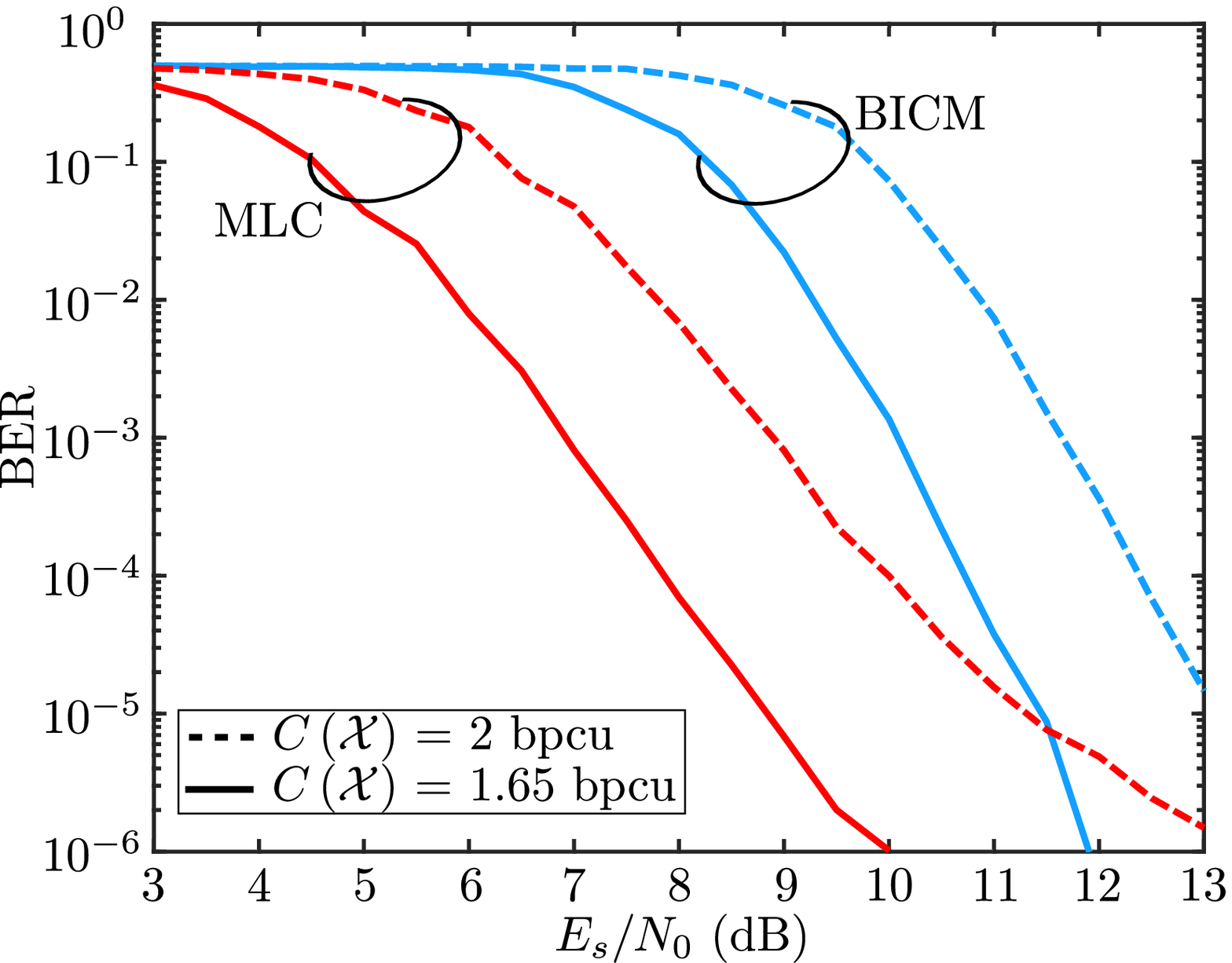}}\hfill{}\subfloat[\label{fig:Fig4(b)} $16\times4$ SSK system]{\centering{}\includegraphics[width=0.5\textwidth]{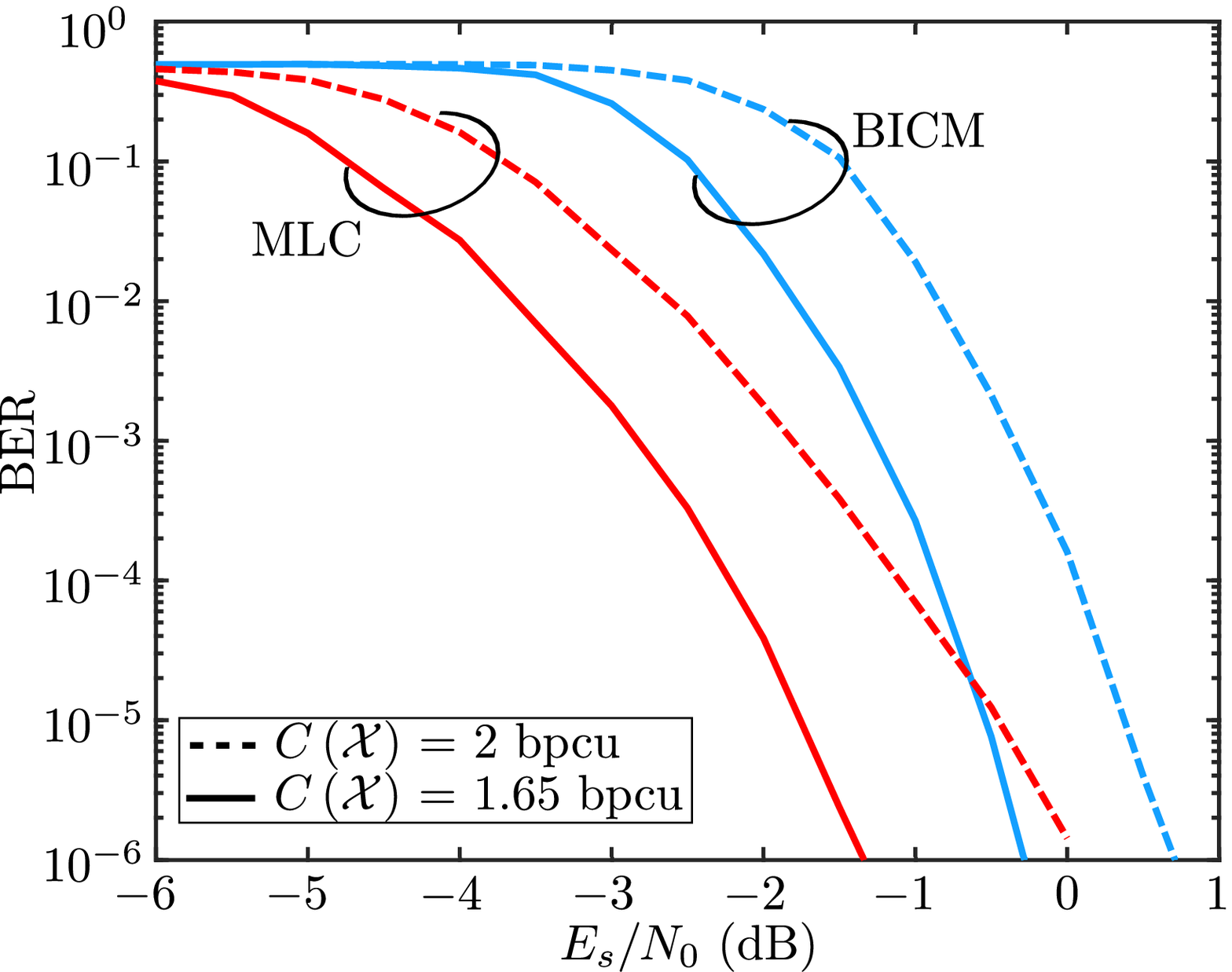}}
\par\end{centering}
\centering{}\caption{\label{fig:Fig4}Bit-error rate curves of BICM and MLC based SSK systems.}
\end{figure*}

\begin{table}[t]
\caption{\label{Tab1}Rates and information bits size of multilevel polar codes}

\centering{}%
\begin{tabular}{|c|c|c|c|c|}
\cline{2-5} \cline{3-5} \cline{4-5} \cline{5-5} 
\multicolumn{1}{c|}{} & \multicolumn{4}{c|}{$C\left(\mathbf{A}\right)$ (bpcu)}\tabularnewline
\cline{2-5} \cline{3-5} \cline{4-5} \cline{5-5} 
\multicolumn{1}{c|}{} & \multicolumn{2}{c|}{$2$} & \multicolumn{2}{c|}{$1.65$}\tabularnewline
\hline 
SSK mode $(N_{T}\times N_{R})$ & $16\times1$ & $16\times4$ & $16\times1$ & $16\times4$\tabularnewline
\hline 
$E_{s}/N_{0}$ (dB) & $3.29$ & $-5.61$ & $1.5$ & $-6.87$\tabularnewline
\hline 
$R_{1}$ & $0.2738$ & $0.3055$ & $0.2037$ & $0.2262$\tabularnewline
\hline 
$R_{2}$ & $0.4143$ & $0.4290$ & $0.3232$ & $0.3346$\tabularnewline
\hline 
$R_{3}$ & $0.5856$ & $0.5671$ & $0.4809$ & $0.4758$\tabularnewline
\hline 
$R_{4}$ & $0.7323$ & $0.6990$ & $0.6456$ & $0.6160$\tabularnewline
\hline 
$K_{1}$ & $70$ & $78$ & $52$ & $58$\tabularnewline
\hline 
$K_{2}$ & $105$ & $110$ & $83$ & $85$\tabularnewline
\hline 
$K_{3}$ & $150$ & $145$ & $123$ & $122$\tabularnewline
\hline 
$K_{4}$ & $187$ & $179$ & $165$ & $158$\tabularnewline
\hline 
\end{tabular}
\end{table}

\section{Results and Discussion\label{sec:Results-and-Discussion}}

We have designed polar codes of length $m_{a}N=1024$ for different
SSK overall capacities. The rates and corresponding information bit
sizes of the component polar codes are shown in Table \ref{Tab1}.
Bit error rate (BER) simulations were run for a maximum of $5\times10^{6}$
frames with a frame error limit of $100$ for all the BER curves. 

To the best of the authors' knowledge, multilevel polar codes have
not previously been designed for use with SSK modulation. Therefore,
the closest benchmark for performance comparison is with the bit-interleaved
polar coded modulation. Fig. \ref{fig:Fig4} shows the BER vs $E_{s}/N_{0}$
curves of BICM and MLC based SSK systems, where $E_{s}/N_{0}$ is
the ratio of the transmitted energy per symbol to the noise power
spectral density. For the $16\times1$ SSK system shown in Fig. \ref{fig:Fig4(a)},
the effective channel is single-input single output (SISO). For $C\left(\mathcal{X}\right)=2$
bpcu, the MLC based design outperforms BICM by $2.3$ dB at a BER
of $10^{-4}$ which diminishes at high SNR $\left(\sim11\right)$
dB and the two curves eventually intersect. On the other hand, the
MLC system with $C\left(\mathcal{X}\right)=1.65$ bpcu outperforms
BICM by $2.9$ dB at a BER$=10^{-4}$ and does not intersect until
a BER of $10^{-6}$.

Fig. \ref{fig:Fig4(b)} shows the BER performance curve for a $16\times4$
SSK system. The MLC based system outperforms BICM by a coding gain
of $1$ dB and $1.5$ dB at a BER of $10^{-4}$ for $C\left(\mathcal{X}\right)=2$
and $1.65$ bpcu, respectively. Here, we again see the same trend
that for high overall SSK capacity, the MLC BER curve approaches the
BICM curve at high SNR. However, the coding gain is lower, as expected,
as compared to the $16\times1$ SSK system because at the receiver
we have four separate received streams, i.e., the effective channel
is SIMO. \balance

\section{Conclusion\label{sec:Conclusion}}

In this paper, we have designed multilevel polar codes for an SSK-modulated
MIMO system. We used the capacity rule to evaluate the bit-level ergodic
capacities of SSK modulation. As the spatial bits choose different
transmitting channels, it is necessary to use Monte Carlo simulation
to find the average bit-level capacities. As the effective channels
are either SISO or SIMO, we base our polar code design on that for
a SISO Rayleigh fading channel. This assumption is not capacity achieving
but provides an easy way to design the multilevel polar coded system
for a MIMO channel. BER simulation results show that the multilevel
polar coded system with multilevel SSK modulation still provides considerable
coding gain compared to the corresponding BICM system with SSK modulation.

This work can also be extended to design of multilevel polar codes
for SM and generalized spatial modulation (GSM), which are suitable
for attaining high spectral as well as energy efficiency.

\appendices{}

\section*{Acknowlegment}

This work was funded by the Irish Research Council under a Consolidator
Laureate Award (grant no. IRCLA/2017/209).

\bibliographystyle{IEEEtran}
\bibliography{references_MLPCSSK}

\begin{IEEEbiography}[{\fbox{\begin{minipage}[t][1.25in]{1in}%
Replace this box by an image with a width of 1\,in and a height of
1.25\,in!%
\end{minipage}}}]{Your Name}
 All about you and the what your interests are.
\end{IEEEbiography}

\begin{IEEEbiographynophoto}{Coauthor}
Same again for the co-author, but without photo
\end{IEEEbiographynophoto}

\end{document}